\documentclass[preprint,nofootinbib,preprintnumbers]{revtex4}
\usepackage{graphicx}
\usepackage{epsfig}
\usepackage{amsmath,amssymb,amsfonts}
\usepackage{color}
\usepackage{amsmath}
\usepackage{amsfonts}
\usepackage{slashed}
\usepackage{framed}
\usepackage{subfig}

\newcommand{\be}{\begin{equation}}
\newcommand{\ee}{\end{equation}}
\newcommand{\ba}{\begin{eqnarray}}
\newcommand{\ea}{\end{eqnarray}}

\newcommand{\bi}{\begin{itemize}}
\newcommand{\ei}{\end{itemize}}

\begin{document}

\preprint{MCTP-11-33}
\title{The Baryon-Dark Matter Ratio Via Moduli Decay After Affleck-Dine
Baryogenesis}
\author{Gordon Kane\footnote{gkane@umich.edu}$^{a}$, Jing Shao\footnote{jishao@syr.edu}$^{b}$, Scott Watson\footnote{gswatson@syr.edu}$^{b}$ and Hai-Bo Yu\footnote{haiboyu@gmail.com}$^{a}$}
\affiliation{\vspace{0.3cm} $^a$Michigan Center for Theoretical Physics, University of
Michigan, Ann Arbor, MI 48109, USA \\
$^b$201 Physics Building, Syracuse University, Syracuse, NY 13244, USA}
\date{\today }

\begin{abstract}
Low-scale supersymmetry breaking in string motivated theories implies the
presence of ${\cal O}(100~{\rm TeV})$ scale moduli, which generically lead to a significant
modification of the history of the universe prior to Big Bang
Nucleosynthesis. Such an approach implies a non-thermal origin for dark
matter resulting from scalar decay, where the lightest supersymmetric particle can account for the observed dark matter relic density.  We study the further effect of the decay on the baryon asymmetry
of the universe, and find that this can satisfactorily address the problem of the over-production
of the baryon asymmetry by the Affleck-Dine mechanism in the MSSM. 
Remarkably, there is a natural connection between the baryon and dark matter abundances today, 
which leads to a solution of the `Cosmic Coincidence Problem'. 
\end{abstract}

\maketitle

\section{Introduction}
Cosmological observations not only determine precisely the relic abundance of dark matter and baryons, but also imply an interesting connection between their relative amounts $ \Omega_{dm} /\Omega_B\approx 5$, leading to what some have called the `Cosmic Coincidence Problem'.  One approach\footnote{There have been many approaches to address the coincidence problem from a single source, but these approaches are typically either involved or require the introduction of large parameters (or both). For example, in \cite{modulidecay} the authors use the decay of a scalar for generating both the baryon asymmetry and the dark matter abundance. This requires a new sector for baryogenesis. In \cite{Qballdecay} Q-ball decay was used to achieve the correct baryon-dark matter ratio through the Affleck-Dine mechanism. Recently, there have been interesting proposals where dark matter is produced at the same time as the baryon asymmetry
via Affleck-Dine mechanism~\cite{otherapproach}. These models also require a new sector for dark matter.} has been to try and realize the origin of both as coming from a single source. In this paper we will take a different approach.

The Minimal Supersymmetric extension of the Standard Model (MSSM) with R-parity has 
all the ingredients to address these issues. The Lightest Supersymmetric Particle (LSP) is a good dark matter candidate that can naturally give rise to the observed dark matter relic density.
Moreover, the existence of many flat directions in the potential with $B-L$ violating operators allows 
the Affleck-Dine~(AD) mechanism to work effectively \cite{Affleck:1984fy}, generating a large baryon asymmetry from scalar decay. However, in this simple MSSM approach, the dark matter density and baryon asymmetry are generated by different mechanisms and at different epochs in the early universe -- they are not correlated in general.  Furthermore, the AD mechanism usually over-produces the baryon asymmetry, resulting in a value which is much higher than the observed value.  In this paper we will argue that by simply accounting for the presence of additional light scalars (moduli) we can resolve these two problems simultaneously.  

Moduli are generically expected from top-down approaches to the MSSM when the theory is UV completed in String/M-theory compactifications. The presence of moduli can significantly change 
the thermal history of the universe \cite{Watson:2009hw}.  In particular, late decays of these fields can interfere with Big Bang Nucleosynthesis leading to a `cosmological moduli problem'.  To avoid this moduli are typically required to have masses of order $10-100$~TeV. These moduli would not only dilute 
the primordial relics but also produce LSP dark matter through their universal gravitational coupling. 
It has been shown in \cite{Moroi:1999zb,Acharya:2008bk,Acharya:2009zt,Acharya:2010af} that non-thermally produced WIMPs from moduli decay can account for the observed dark matter abundance. On the other hand, the entropy production from the moduli automatically provides a mechanism to reduce the overproduced baryon asymmetry from AD mechanism. In this paper, we consider this approach under conditions where non-thermal production provides the right dark matter abundance, 
and ask if the observed baryon asymmetry can be simultaneously achieved. 

Previous suggestions for using the decay of scalars to address the over-production of the 
baryon asymmetry in AD baryogenesis appeared in \cite{Affleck:1984fy,Ellis:1987rw,Campbell:1998yi, Dolgov:2002vf,Barenboim:2008zk}.  Here we realize this idea for the first time in a fundamental theory, where tight constraints may be placed on the underlying parameters.  Moreover, using this approach we find a natural explanation
for the relationship between the amount of baryon and dark matter because they result from moduli decay.
These moduli and other scalars including sfermions
have masses $m_{i}\simeq m_{3/2}\sim {\cal O}(50)$~TeV. 
This is a generic consequence of SUSY theories with heavy scalars which are required to not only yield
realistic low-energy phenomenology (give rise to electroweak symmetry breaking and generate hierarchies), but also
be consistent (e.g. anomaly-free) at high energies and in the presence of gravity~\cite{hierarchy}. This result is
independent of the details of SUSY breaking and very difficult to evade, as
was recently discussed for the case of gauge mediation in \cite{Fan:2011ua}.

We now summarize our main conclusions. 
We find that acceptable values of the baryon asymmetry can 
be realized from the combination of entropy from moduli decay and a large initial baryon asymmetry 
as naturally arises from the AD mechanism in the MSSM.   We also find that for the same expected values associated with the moduli decay the correct dark matter abundance can result. We note that both the
baryon asymmetry and dark matter abundance are essentially determined
by the reheat temperature and the mass of the scalar, and this gives a new explanation for the `cosmic coincidence problem'.

In the next section we briefly review the AD mechanism for baryogenesis. 
Next we turn to the late-time production of entropy associated with the decay of moduli and demonstrate how this can lead to acceptable values not only for the baryon asymmetry and dark matter density, but also offers an explanation for the relative abundance today.  We then summarize with our conclusions.

\section{A Brief Review of Affleck-Dine Baryogenesis}
In this section we briefly review the AD mechanism of baryogenesis. 
For a more detailed review with references to the original literature we refer the reader to \cite{Dine:2003ax}.
The AD mechanism is realized through the existence of the many approximately flat directions in the MSSM -- which arise from products of squark and slepton fields. These flat directions are expected to be lifted by non-renomalizable operators and the corresponding scalar fields (AD fields) then develop large Vacuum Expectation Values (VEVs) in the early universe.  These VEVs may break baryon or lepton number, and non-zero CP-violating phases can result from SUSY breaking effects.  The final Sakharov condition for baryogenesis is then met by the expansion of the universe, which provides the out-equilibrium condition necessary to generate the net baryon asymmetry.  

The relevant potential for the AD field $\phi$ in the early Universe is \cite{Dine:1995uk,Dine:1995kz}
\begin{equation}
V(\phi )=(-cH^{2}+m_{\phi}^{2})|\phi |^{2}+\left( \frac{aH+Am_{3/2}}{M^{n-3}%
}\lambda \phi ^{n}+h.c.\right) +|\lambda |^{2}\frac{|\phi |^{2n-2}}{M^{2n-6}}%
,\label{vphi}
\end{equation}%
where $c$, $a$, $A$ and $\lambda$ are order one constants. 
The origin of the terms in the potential are easy to understand. In the early universe
the gravitational background or the presence of finite temperature will break SUSY, e.g. during inflation. 
This leads to a Hubble-scale mass and Hubble-scale A-terms for the AD field.  At lower energy scales 
SUSY breaking soft terms become dominant and generate a soft mass for the AD field ($%
m_{\phi }$) and additional A-terms, which are of order the gravitino mass $%
m_{3/2}$. The last term in the potential corresponds to a higher dimensional operator 
in the superpotential, $W \supset \lambda \phi^{n}/M^{n-3}$, which acts to lift the flat direction. Here $M$ is 
the cutoff scale where new physics appears and is naturally expected to be near the GUT or 
reduced Planck scale $M_{p}\simeq 2.4\times 10^{18}$ GeV. 
It was pointed out in~\cite{Dine:1995uk,Dine:1995kz} that the Hubble mass term (\ref{vphi}) must be tachyonic
for successful AD baryogenesis and we will adapt this standard assumption throughout the remainder of this paper\footnote{For a recent study of the behavior of the AD field during and following inflation see \cite{DuttaMarsh}.}.

The AD field exhibits different behavior depending on the cosmological epoch in which we consider. During
high-scale inflation $H \gtrsim m_{3/2}$ and the soft terms in~(\ref{vphi}) are
negligible. The AD field will then have a Hubble-scale mass and so is
almost critically damped, relaxing into small oscillations about its minimum
within a few efoldings irrespective of its initial displacement. The minimum of the potential during this epoch is given by
\begin{equation}
\langle \phi \rangle \sim M\left( \frac{H}{M}\right) ^{1/(n-2)}.
\label{phi00}
\end{equation}%

As the expansion rate decreases $H
$ will eventually become comparable to $m_{3/2}$, and the Hubble induced
terms in~(\ref{vphi}) become of the same order as the soft breaking terms.
The AD field then begins large oscillations when $H \lesssim m_{\phi }\sim m_{3/2}
$, forming a scalar condensate which evolves as non-relativistic matter. It
is in this period that there exist both CP violation and baryon number
violation, and non-zero baryon number is generated in the AD condensate in
the usual way \cite{Dine:2003ax}. The baryon number generated at this epoch is given by
\begin{eqnarray}
n_B\simeq\frac{1}{M^{n-3}}\sin(\delta)\phi^n_0,
\end{eqnarray}
where $\delta$ is the CP-violating phase, and $\phi_0$ is the VEV of the AD field at $H\sim m_{\phi}$. Using (\ref{phi00}) and that during oscillations we have $H\sim m_{\phi}$ we find the VEV 
\begin{eqnarray}
\phi_0\sim M\left(\frac{m_\phi}{M}\right) ^{1/(n-2)}.
\label{phi0}
\end{eqnarray}
The ratio of baryon number density to AD field at this epoch is $\left({n_B}/{n_\phi}\right)_i\simeq\sin\delta,$ where $n_\phi\simeq m_\phi\phi^2_0$. Note that $(n_B/n_\phi)_i$ depends on the CP-violating phase and can be as large as ${\cal O}(1)$.  We note that these are phases during the AD oscillations and not related to the phases of the soft SUSY breaking Lagrangian.

When the Hubble expansion rate becomes much less than $m_{3/2}$ the baryon number of the condensate is frozen-in, and later will be converted into the baryon asymmetry.  The inflaton decays around the time scale $\sim\Gamma^{-1}_I$, where $\Gamma_I\sim m^3_I/m^3_p\simeq10^{9}~{\rm GeV}$ for $m_I\sim 10^{12}~{\rm GeV}$. The baryon number density at this epoch is 
\begin{equation}
n_{B}(t\sim\Gamma _{I}^{-1})\sim m_{\phi }\phi _{0}^{2}\left( \frac{\Gamma _{I}%
}{m_{\phi }}\right) ^{2}\left( \frac{n_{B}}{n_{\phi }}\right) _{i},
\end{equation}
where $\Gamma _{I}/m_{\phi }$ comes from the expansion of the universe. After the inflaton decay, the inflaton energy is converted to radiation where the reheating temperature is $T^I_R\sim\sqrt{\Gamma_I M_p}\simeq~10^{9}~{\rm GeV}$ and the photon density is given by $n_\gamma\sim {T^I_R}^3$.  Therefore, using (\ref{phi0}) for the value of $\phi_0$ we find the baryon to photon ratio
\begin{eqnarray}
\frac{n_{B}}{n_{\gamma }} \sim \frac{T_{R}^{I}}{m_{\phi }}\frac{\phi
_{0}^{2}}{M_{p}^{2}}\left( \frac{n_{B}}{n_{\phi }}\right) _{i}\sim \frac{T_{R}^{I}M}{M_{p}^{2}}\left( \frac{M}{m_{\phi }}\right) ^{\frac{n-4}{n-2}}\left( \frac{n_{B}}{n_{\phi }}\right) _{i}.
\label{initial_baryon_photon_ratio}
\end{eqnarray}

From~(\ref{initial_baryon_photon_ratio}) we see that for $n=4$, $(n_B/n_\phi)_i\sim 1$ and $M\sim M_{p}$ this is within the correct range to explain the observed baryon asymmetry if there is no significant late-time entropy production, i.e. in an approach that does not account for the presence of moduli.
For larger $n\gg 1$ the scalar initial VEV $\phi _{0}$ can be as large as $M
$, resulting in significantly larger baryon asymmetry. For example, in the MSSM
the flattest direction requires an operator with $n=9$ to lift
it~{\cite{Gherghetta:1995dv}}. This indicates that for this particular flat direction decay would result in
a baryon to photon ratio $n_{B}/n_{\gamma }\sim 2$ for $M\sim M_{p}\sim10^{18}$~GeV, or $n_{B}/n_{\gamma }\sim 10^{-4}$ for $M\sim M_{\mathrm{GUT}}\sim
10^{16}$~GeV.  
Therefore, the baryon asymmetry is typically over produced from the AD mechanism in the MSSM. This points
toward models with large entropy production at late times from moduli decay.

\section{Late-time entropy production and dark matter genesis}
\subsection{Baryon asymmetry after moduli decay}
\label{baryonafter}

Now let us consider a simple case with one modulus $X$ decaying long after the
AD field decayed to see how to estimate the needed numbers. The evolution of 
moduli after inflation is similar to that of the AD flat directions discussed in the previous section. 
However, since moduli originate from the coordinates of compact extra dimensions, 
they have a quite different potential from that of the AD flat direction. 
Generically, it is expected to have all renormalizable terms present in the potential. 
In the early universe with large inflaton energy density, these terms receive large Hubble corrections. 
This typically leads to a Planck scale displacement for moduli fields from their low-energy minimum $X_0 \sim M_p$~\cite{Dine:1995kz}. 

Since the modulus couples gravitationally to all MSSM particles it generically decays to SM particles
and their superpartners with branching fractions of the same order of
magnitude. 
The rest of the decay goes to SM particles which are then thermalized, resulting in a
significant increase in the total entropy. The decay width of the modulus
can be parameterized as 
\begin{equation}
\Gamma _{X}=D_{X}\frac{m_{X}^{3}}{M_{p}^{2}}
\end{equation}%
where $M_{p}$ is the reduced Planck scale and $D_{X}$ is a constant determined
by the moduli to matter couplings and typically takes values of ${\cal O}(1)$ in estimates arising from string compactifications \cite{Acharya:2008bk}.

Given the large initial displacement of the moduli field and its long lifetime it will come to dominate the 
energy density of the universe prior to its decay. The ratio of the moduli number density to entropy density before the moduli decay is
determined by the initial moduli amplitude and the reheating temperature in
a similar way as the baryon asymmetry 
\begin{equation}
\label{eqnsc}
Y_{X}^{0}\equiv \frac{n_{X}}{s}\simeq \frac{3}{4}\frac{T_{R}^{I}}{m_{X}}%
\left( \frac{X_{0}}{M_{p}}\right) ^{2}
\end{equation}

where $X_{0}$ is the amplitude at the start of the moduli oscillation, and
we use an upper index ${}^{0}$ to distinguish the yield after the modulus
decay. Compared to (\ref{initial_baryon_photon_ratio}), we can see that
the baryon to moduli ratio is determined by the
initial amplitudes and masses of the fields
\begin{equation}
\frac{Y_{B}^{0}}{Y_{X}^{0}}\simeq \left( \frac{m_{X}}{m_{\phi }}\right)
\left( \frac{\phi _{0}}{X_{0}}\right) ^{2} \left( \frac{n_{B}}{n_{\phi }}\right) _{i} .  \label{ratio}
\end{equation}%
Since this ratio is unaffected by the moduli decay (it is a comoving quantity and so does not depend on the expansion) 
it can be used to determine the baryon number density after moduli decay,
\begin{equation}
Y_{B}^{0} \rightarrow Y_B=\frac{Y_{B}^{0}}{\Delta} = \frac{n_B}{s_{\text{after}}} \simeq \frac{n_{X}}{s_{\text{after}}}\left(\frac{Y_B^0}{Y_{X}^0}\right) \simeq \frac{3}{4}\frac{T_{R}^{X}}{m_{\phi }}\left( \frac{\phi _{0}}{X_{0}}\right) ^{2} \left( \frac{n_{B}}{n_{\phi }}\right) _{i},
\end{equation}%
where $\Delta=s_{after}/s_{before}$ is the dilution from decay and we have made use of (\ref{eqnsc}).
Here $n_B$ and $n_X$ are the number densities of baryons and moduli at the time of decay and $s_{\text{after}}$ is the entropy density after the decay. The $Y_{B}$ obtained above is related to the baryon to photon ratio today
given by the equation $n_{B}/n_{\gamma }\simeq 7.04\;Y_{B}$. Here the 
factor $7.04$ is the entropy to photon ratio at the current epoch. 
Then the baryon to photon ratio today is 
\begin{equation}
\frac{n_{B}}{n_{\gamma }}\simeq 4.5\times 10^{-10}\times \left( \frac{%
T_{R}^{X}}{64\;\text{MeV}}\right) \left( \frac{75\;\text{TeV}}{m_{\phi }}%
\right) \left( \frac{\phi _{0}/X_{0}}{10^{-2}}\right) ^{2}
\end{equation}
where we have taken $(n_B/n_{\phi})_{i}\sim 1$ and we have chosen fiducial values which 
are typical from the underlying theory and can simultaneously yield the correct abundance of dark matter: $%
D_{X}=4$, $m_{X}\simeq 2m_{3/2}=150$~TeV. The resulting reheat
temperature is given by $T_{R}^{X}\simeq (90/\pi ^{2}g_{\ast })^{1/4}(\Gamma
_{X}M_{p})^{1/2}\simeq 64$~MeV, where $g_{\ast }\simeq 15$ was used.
For $\phi_0/X_0\sim 10^{-2}$, the obtained ratio is just the right number to compare with the
observed asymmetry $n_{B}/n_{\gamma }=6.1\times 10^{-10}$. Note that because the reheat temperature $T_R^{X} \propto m_{X}^{3/2}\sim m_{\phi}^{3/2}$, there is only a mild dependence on $m_{\phi}^{1/2}$. 

The above result shows that the baryon to photon ratio in this approach is intimately related to the ratio of the initial amplitudes of the AD field and the modulus, $\phi_0/X_0$. This is easy to understand since the photon density is dominantly generated from the modulus decay. As we have discussed in Section II, the initial amplitude for the AD field is calculable and is given in (\ref{phi0}). Note that $\phi_0$ depends nontrivially on the dimension of the non-renormalizable operator that lifts the flat direction. Since larger $n$ leads to larger $\phi_0$ and therefore larger contribution to the baryon asymmetry, we can focus on the flattest directions in MSSM that require the largest $n$ to get lifted\footnote{We assume all non-normalizable operators that are allowed by gauge invariance and R-parity are generated.}. As showed in Ref.~{\cite{Gherghetta:1995dv}}, the flattest direction (one of the $Q, u, e$ combinations) corresponds to $n=9$. 
Assuming that the non-renormalizable operator is generated at the reduced Planck scale $M\sim M_p$ 
and taking $m_{\phi}\sim 10^5$~GeV we find $\phi_0\sim 10^{16}$~GeV. 
For the next flattest direction (one of the $d,L$ combinations) -- which is not lifted until $n=7$ --
we have $\phi_0\sim 3\times 10^{15}$~GeV. So we can see that these flattest directions in the MSSM naturally 
have amplitudes two or three order of magnitudes smaller than $M_p$, i.e., $\phi_0/X_0\sim 10^{-3}-10^{-2}$. This ``little hierarchy" is exactly what is needed to explain the baryon asymmetry observed. Its origin can be traced back to the matter content and gauge structure of the MSSM. 

\subsection{Dark matter density}
\label{ratioafter}
As discussed in subsection \ref{baryonafter} moduli decay to superpartners
with a large branching ratio. Each of these superpartners will eventually decay
to an LSP and so typically there are $2B(X\rightarrow \chi \chi )$ LSPs produced
per moduli, where $B(X\rightarrow \chi \chi )$ is the branching fraction
for moduli decay to superpartners. Therefore, the yield of LSPs after the
decay is given by 
\begin{equation}
Y_{\chi }=2B(X\rightarrow \chi \chi )Y_{X}=\frac{3}{2}B(X\rightarrow \chi \chi)\frac{%
T_{R}^{X}}{m_{X}}.
\end{equation}%
The produced LSPs undergo an out-of-equilibrium annihilation. For that to
occur the self annihilation rate must be
larger than the expansion rate $n_{\chi }\langle \sigma v\rangle >H$, which leads to the following condition 
\begin{equation}
n_{\chi } \gtrsim n_{\chi }^{c} \simeq \frac{H}{\langle \sigma v\rangle }\bigg |%
_{T=T_{R}^{X}}
\end{equation}%
where $n_{\chi }^{c}$ is the critical density for annihilations. For $%
T_{R}^{X}\approx 100$~MeV and $m_{X}\approx 10^{5}$~GeV we find that
the abundance is too large ($Y_{\chi }\approx 10^{-7}$ vs $Y_{\chi }^{c}\approx
10^{-11}$ ) and LSPs will further annihilate. The final abundance is
determined by the critical number density $n_{\chi }^{c}$ from the
out-of-equilibrium annihilation of LSPs. The final dark matter yield is
\begin{eqnarray}
Y_{\chi }\simeq \frac{n_{\chi }^{c}}{s} &\simeq&\frac{45}{2\pi ^{2}g_{\ast }}%
\frac{H}{T^{3}\langle \sigma v\rangle }\bigg |_{T=T_{R}^{X}} 
\notag \\
&\simeq&\frac{1}{4}\left( \frac{90}{\pi ^{2}g_{\ast }}\right) ^{1/2}\frac{1}{%
M_{p}T_{R}^{X}\langle \sigma v\rangle }
\end{eqnarray}%
The above equation can be converted into the relic abundance today, 
\begin{equation*}
\Omega _{LSP}=\frac{m_{LSP}Y_{\chi }}{\rho _{c}/s_{0}}\simeq
0.11h^{-2}\times \left( \frac{m_{\chi }}{100\;\text{GeV}}\right) \left( 
\frac{3\times 10^{-7}\;\text{GeV}^{-2}}{\langle \sigma v\rangle }\right)
\left( \frac{64\;\text{MeV}}{T_{R}^{X}}\right), 
\end{equation*}%
where $\rho _{c}$ and $s_{0}$ are the current critical density and entropy density, and
their ratio is given by $\rho _{c}/s_{0}\simeq 3.6\times 10^{-9}\;h^{2}\;%
\text{GeV}$. For the non-thermal history that we are considering if a neutralino is to be the dark matter candidate it must be primarily wino-like meaning a larger annihilation cross section which is given by~\cite{Moroi:1999zb}
\begin{equation*}
\langle \sigma v\rangle =\frac{g_{2}^{4}}{2\pi }\frac{1}{m_{\chi }^{2}}\frac{%
(1-x_{w})^{3/2}}{(2-x_{w})^{2}},
\end{equation*}%
where $g_{2}\simeq 0.65$, $x_{w}=m_{W}^{2}/m_{\chi }^{2}$ with $m_{W}\simeq
80.4~\mathrm{GeV}$. For $m_{\chi }=100$~GeV, the annihilation rate is $%
3.3\times 10^{-7}\;\text{GeV}^{-2}$.

Finally the baryon to dark matter ratio today is
\begin{equation}
\frac{\Omega _{B}}{\Omega _{\chi }}\simeq 0.2\times \left( \frac{100\;\text{%
GeV}}{m_{\chi }}\right) \left( \frac{T_{R}^{X}}{64\;\text{MeV}}\right)
^{2}\left( \frac{\langle \sigma v\rangle }{3\times 10^{-7}\;\text{GeV}^{-2}}%
\right) \left( \frac{75\;\text{TeV}}{m_{\phi }}\right) \left( \frac{\phi
_{0}/X_{0}}{10^{-2}}\right) ^{2}\label{baryonDMratio}
\end{equation}%
 This approach can naturally reproduce nearly the observed
baryon-to-dark matter ratio today. It is easy to understand each 
of the relevant factors in (\ref{baryonDMratio}). The dependence on the moduli reheat temperature $T_{R}^{X}$
follows because higher values increase the baryon asymmetry since the moduli density was then higher in the early universe. Moreover, higher values will decrease the dark matter density because of the corresponding increase in entropy production at the time of decay.  The dependence on the averaged annihilation cross-section and velocity
$\langle \sigma v\rangle $ is understood because the amount of dark matter depends inversely
on its ability to self annihilate. The dependence on the AD field $1/m_{\phi }$ comes from the 
number density of the AD field (flat direction) $n_{\phi }\sim \rho _{\phi }/m_{\phi }$. In
fact, as mentioned in the beginning of Section II, the AD mass $m_{\phi }$ is
of the same order as the gravitino and moduli mass $\sim m_{3/2} \sim m_{X}/2$.  
Thus, the true dependence of the
baryon to dark matter ratio on the mass scale is $\sim m_{3/2}^{2}$ after
rewriting the reheating temperature in terms of the moduli mass. In the last
factor, $\phi _{0}$ and $X_{0}$ are the initial amplitudes of the AD flat
direction and the modulus. The factor arises from the ratio of their
corresponding energy densities and determines how much baryon asymmetry is
left after the dilution.

Our results are derived assuming that the AD condensate evolved homogeneously after it formed. 
In general, it is also possible that the AD condensate becomes unstable with respect to spatial perturbations and turns into non-topological solitons, so-called Q-balls~\cite{Qball}.
In such a case, Q-balls can decay very late and greatly change the resulting baryon asymmetry. 
Nevertheless, as we have checked, in the approach considered here, where gaugino 
masses are suppressed compared to the scalar mass, a large set of flat directions with second and third 
generation squarks will not fragment into Q-balls, in contrast to the more usual result based on the
MSSM spectrum with one mass scale. This includes the flattest directions that 
are lifted at the level $n=7$ and $n=9$. The associated condensates if formed are likely to 
dominate the energy density compare to all other flat directions. 
If Q-balls do form from other less flat directions, they will likely decay before the moduli decay and their contribution will be washed away.  A full detailed treatment of the Q-ball in our approach is beyond the scope of this paper, and 
will appear elsewhere. 

\section{Conclusions}

In this paper, we have studied AD baryogenesis and the baryon-dark matter so-called
coincidence problem in the MSSM accounting for the presence of moduli and the possibility of a non-thermal history for the early universe.   Such an approach emerges from String/M theory compactifications with stabilized moduli and realistic soft supersymmetry breaking. 
For such an approach, it is natural for the baryon asymmetry to arise via 
the AD mechanism in which MSSM flat directions with $U(1)_{B-L}$ charge
form a condensate which later decays into baryons. 
In many instances this mechanism is too efficient and gives a baryon asymmetry of 
order unity. However, here we have seen that when moduli decay shortly before BBN, the resulting entropy dilution leads to an acceptable baryon asymmetry. As discussed in detail in the text, for moduli and gravitino masses of order $100$~TeV and reheat temperature of order $100$~MeV, the resulting baryon to dark matter ratio is  
\be
\frac{\Omega _{B}}{\Omega _{\chi }}\simeq 0.2\times \left( \frac{100\;\text{%
GeV}}{m_{\chi }}\right) \left( \frac{T_{R}^{X}}{64\;\text{MeV}}\right)
^{2}\left( \frac{\langle \sigma v\rangle }{3\times 10^{-7}\;\text{GeV}^{-2}}%
\right) \left( \frac{75\;\text{TeV}}{m_{\phi }}\right) \left( \frac{\phi
_{0}/X_{0}}{10^{-2}}\right) ^{2}
\ee
implying a fundamental relation between the amounts of baryonic and dark matter.
Moreover, for the same set of parameters the dark matter abundance is in near agreement with cosmological observations.
We emphasize that these results are robust and hold in a large class of string compactifications with stabilized moduli.  They do not require the addition of ad-hoc or special mechanism.

\section*{Acknowledgments}
We would like to thank M. Cvetic, D. Feldman, H. Murayama, R. Penco and K. Sinha for useful discussions. HY thanks Aspen Physic
Institute for hospitality where part of work was done. 
J.S. and S.W. would like to thank Cornell University for hospitality and are both supported by Syracuse University College of Arts and Sciences.

\appendix

\end{document}